\documentclass[10pt,psfig,epsfig,bbox]{article}  
\begin{document}  
\newcommand{\vk}{{\vec k}}  
\newcommand{\vK}{{\vec K}}   
\newcommand{\vb}{{\vec b}}   
\newcommand{{\vp}}{{\vecp}}   
\newcommand{{\vq}}{{\vec q}}   
\newcommand{\vQ}{{\vec Q}}  
\newcommand{\vx}{{\vec x}}  
\newcommand{\vh}{{\hat{v}}}  
\newcommand{\tr}{{{\rm Tr}}}   
\newcommand{\beq}{\begin{equation}}  
\newcommand{\eeq}[1]{\label{#1} \end{equation}}   
\newcommand{\half}{{\textstyle\frac{1}{2}}}   
\newcommand{\lton}{\mathrel{\lower.9ex \hbox{$\stackrel{\displaystyle  
<}{\sim}$}}}   
\newcommand{\ee}{\end{equation}}  
\newcommand{\ben}{\begin{enumerate}}   
\newcommand{\een}{\end{enumerate}}  
\newcommand{\bit}{\begin{itemize}}   
\newcommand{\eit}{\end{itemize}}  
\newcommand{\bc}{\begin{center}}   
\newcommand{\ec}{\end{center}}  
\newcommand{\bea}{\begin{eqnarray}}   
\newcommand{\eea}{\end{eqnarray}}  
\newcommand{\beqar}{\begin{eqnarray}}   
\newcommand{\eeqar}[1]{\label{#1}  
\end{eqnarray}}   
\newcommand{\gton}{\mathrel{\lower.5ex \hbox{$\stackrel{> } 
 {\scriptstyle \sim}$}}} 
 
\vspace*{.4in}  
\begin{center} 
{\LARGE \bf Anomalous Anti-proton to  Negative Pion Ratio  \\[.ex]  
as Revealed by Jet Quenching at RHIC \\ 
} 
\vspace*{0.3in}   
{\Large Ivan~Vitev$^{*}$ and Miklos~Gyulassy$^{*\,\dagger}$}  
\vspace*{0.25in}   
  
{\it $^*$  Department of Physics, Columbia University,   
         538 W. 120-th Street, New York, NY 10027, USA\\[.5ex]  
         $^\dagger$ Collegium Budapest, Szentharomsag u.2 
H-1014 Budapest, Hungary  }  
\vspace*{0.2in}

\begin{center} 
\begin{minipage}[t]{15.4cm} 
        {\small  \noindent    
{\bf Abstract.} We study the apparent  
discrepancy  between the  standard PQCD predictions  
for the meson and baryon ratios and multiplicities  
at moderate high $p_{\rm T} > 2$~GeV and recent  
experimental measurements in $Au+Au$ collisions at 
$\sqrt{s}_{NN}=130$~GeV at the  
Relativistic Heavy Ion Collider (RHIC).  
We show that the differences, most pronounced  
in central collisions, can be explained by a    
strong non-perturbative baryon Junction component, 
which dominates the currently accessible experimental $p_{\rm T}$ window  
and the non-abelian energy loss of fast partons propagating through  
hot and dense medium.  
The recently introduced two component hybrid model, which combines  
a quenched jet PQCD  calculation in the Gyulassy-Levai-Vitev (GLV)  
formalism and a phenomenological ``soft" part, is further elaborated to  
take into account      
the full 3D expansion in the pre-hadronization phase and include 
particle flavor dependent ``soft" inverse slopes as suggested by the 
baryon Junction picture. 
We show that such approach can resolve what seems to be a factor of  
 $\simeq 2$ difference in the moderate high $p_{\rm T}$ suppression  
of $\pi^0$  and  $h^-$ as recently reported by the PHENIX collaboration.      
The observed  quenching of the high $p_{\rm T}$  particle spectra and the  
large $\bar{p}/\pi^-$ and $p/\pi^+$ ratios as a function of $p_{\rm T}$  
are found to be consistent with a creation of a deconfined phase and  
non-abelian energy loss of fast partons in a plasma of initial gluon rapidity 
density $dN^g/dy \sim 1000$ 
}

\end{minipage} 
\end{center} 
 
\end{center} 
 
\vspace*{0.25in}

 
\begin{center} 
{\large  \bf INTRODUCTION} 
\end{center}

The preliminary data from the first $Au+Au$ run of the Relativistic  
Heavy Ion Collider (RHIC) at $\sqrt{s}_{NN}=130$~GeV was reported  
by the experimental collaborations (BRAHMS, PHENIX, PHOBOS and STAR) 
at  the Quark Matter 2001~[1] conference.  
It suggests  the observation of qualitatively new  phenomena in comparison to  
the CERN Super Proton Synchrotron (SPS)  $Pb+Pb$ program results at   
$\sqrt{s}_{NN}=17.4$~GeV. 
First, the moderate high $p_{\rm T}$ spectra of neutral pions  in the $10\%$  
central collisions  were reported by PHENIX~[2]  
to be suppressed by a factor 2-4 relative to the PQCD predictions scaled by 
nuclear geometry ($T_{AB}({\bf b})$, the Glauber profile density   
per unit area at an impact parameter ${\bf b}$). 
Second, the analysis of semi-peripheral collisions   
by STAR~[3] 
shows that the elliptic flow $v_2(p_{\rm T})$  
is compatible with  the monotonically growing hydrodynamic predictions   
for small transverse momenta, but saturates at $\sim 0.15$  
above $p_{\rm T} >  
2\;{\rm GeV}$. Third,  studies of the charged particle rapidity 
density as a function of centrality by PHOBOS and  
PHENIX~[4] revealed the  
expected (but previously not  observed)  $T_{AB}({\bf b})\sim  
N_{part}^{4/3}$ mini-jet  
component~[5]. 
Finally, in the $p_{\rm T} = 1-3.5$~GeV  
 window where both $\pi^0$ and  
$h^-$ have been measured, the $\pi^0$ are quenched by a factor 
of 2 more than the negative hadrons.

While the  data are preliminary, the analysis of   
Ref.~[6] suggests 
that these first results from RHIC may  be interpreted as  
 evidence for the production of a dense gluon plasma   
in central and semi-central reactions. The key diagnostic tool for 
probing the properties of the medium in that analysis 
is jet quenching~[7]. Estimates  
based on   high $p_{\rm T}$  suppression  of negative hadrons indicate  
that the initial gluon density may have reached  
$\rho_g(\tau\sim 0.2\; {\rm fm/c}) \sim 100 \rho_0$,  
where $\rho_0=0.15\;{\rm fm}^{-3}$ is  the nuclear saturation density.   
At RHIC energies, $\sqrt{s}_{NN} = 100-200$~GeV,  the initial flux of high  
$p_{\rm T}$ jets becomes   
sufficiently large and the produced gluon plasma is sufficiently dense 
and long lived to make jet quenching finally observable. 
In contrast, at SPS energies the  semi-hard $\pi^0$ spectrum 
reported by the WA98 collaboration~[8] was 
found to be actually  
{\em enhanced} by a factor of 2 at  $p_{\rm T} \sim 2-4$~GeV  
due to the Cronin effect~[9]. 
We emphasize the importance  of studying  $p+A$ systems 
to resolve the strength 
of the Cronin versus nuclear shadowing effects on the 
high $p_{\rm T}$ yields.

An unexpected result at RHIC  
reported by  PHENIX~[10]
was that the  
flavor composition of high $p_{\rm T} > 2$ GeV 
 negative and positive $p_{\rm T}$  differential hadron yields   
may actually be dominated by anti-protons and protons respectively.  
Such behavior has not been observed so far in  $pp$ or $AA$ collisions.  
In the conventional QCD multi-particle production phenomenology  
high $p_{\rm T}$ hadrons arise  from the fragmentation  
of colored partons, ($q,g$), whose spectra are computable via PQCD.  
If hot and dense matter is created in $AA$ reactions,  
those partons naturally lose energy through elastic and radiative  
multi-particle final state interactions. The quenching of high $p_{\rm T}$  
partons is thus expected to result in the  quenching of {\em all}  their  
hadronic fragments including $\bar{p}$ and $p$.    
 
To resolve the new 
high $p_{\rm T}$ baryon puzzle at RHIC we propose that the  
non-perturbative baryon  
Junctions component~[11]  
of hadronic spectra  provides a natural mechanism that could account  
for the anomalously large baryon/meson ratios.  This component 
competes  with the semi-hard PQCD processes even in $pp$ and $pA$ 
reactions but has only now been exposed in central 
$Au+Au$ at RHIC, because the usual 
perturbative QCD pion and kaon component   
is strongly  suppressed due to jet quenching.  
The Junction mechanism is further supported by the observation that 
in central collisions at midrapdity $\bar{p}/p\approx 0.65$ and  
$dN^B/dy \sim dN^p/dy-dN^{\bar{p}}/dy \sim 15$~[12],
indicating a very significant baryon number transport (stopping) 
five units of rapidity from the fragmentation regions. 
 
\vspace*{4mm} 
 
\begin{center} 
{\large  \bf HARD AND SOFT HADRONIC COMPONENTS} 
\end{center}

The  PQCD approach  
expresses the differential hadron cross section in $p+p \rightarrow h+X$  
through a convolution of the measured structure functions  
$f(x_\alpha,Q_\alpha^2)_{\alpha/p}$  for the interacting partons  
($\alpha = a,b$),  with the parameterized fragmentation function  
$D_{h/c}(z,Q^2_c)$ for 
the leading scattered parton $c$ into a hadron of flavor $h$ and the 
elementary parton-parton cross sections  
$d\sigma^{(ab \rightarrow cd)}/d\hat{t}$. Next to leading 
order corrections are 
accounted for by the constant $K\sim 2$. Other degrees of freedom 
for PQCD fits to data include the choice of the renormalization 
scale $Q^2$ and    
the  parton broadening  due to initial state radiation, which    
can  be included through a normalized  $k_{\rm T}$-smearing distribution.  
A simple  functional form is the Gaussian smearing 
$f(k_{\rm T})=e^{-k_{\rm T}^2/\langle k_{\rm T}^2 \rangle}/ 
\pi\langle k_{\rm T}^2 \rangle $, 
where $\langle k_{\rm T}^2 \rangle \simeq 0.6-1.0$~GeV$^2$ in $pp,\;  
\bar{p}p$ collisions.  
In $AA$ reactions one expects larger initial parton broadening due to  
 $k_{\rm T}$ kicks.  
In the absence of a medium, the invariant hadron inclusive cross section  
is  given by 
\begin{eqnarray} 
E_{h}\frac{d\sigma_h^{pp}}{d^3p} &=& 
K   \sum_{abcd} \int\! dz_c dx_a  
dx_b \int d^2{\bf k}_{{\rm T}a} d^2{\bf k}_{{\rm T}b}  
 f({\bf k}_{{\rm T}a})f({\bf k}_{{\rm T}b}) 
f_{a/p}(x_a,Q^2_a) f_{b/p}(x_b,Q^2_b) \nonumber \\[1ex] 
&\;&  D_{h/c}(z_c,{Q}_c^2)  
 \frac{\hat{s}}{\pi z^2_c} \frac{d\sigma^{(ab\rightarrow cd)}} 
{d{\hat t}} \delta(\hat{s}+\hat{u}+\hat{t}) \; , 
\label{hcrossec} 
\end{eqnarray} 
where $x_a, x_b$ are the initial momentum fractions carried  
by the interacting partons, $z_c=p_h/p_c$ is the momentum fraction carried  
by the observed hadron.  
 
In the case of jet production inside a hot and dense medium, the fast parton 
loses a fraction of its energy due to gluon radiation induced by final 
state interactions. The full solution for the gluon radiative 
double differential distributions  for the case of  
finite jet energies and few collisions  was found in~[7]
in a form easily integrated numerically  (to $\Delta E$)
taking into account both the kinematic bounds and the running coupling  
$\alpha_s({Q}^2)$.  
 
\newpage 
 
\vspace*{7.5cm} 
\begin{center} 
\includegraphics{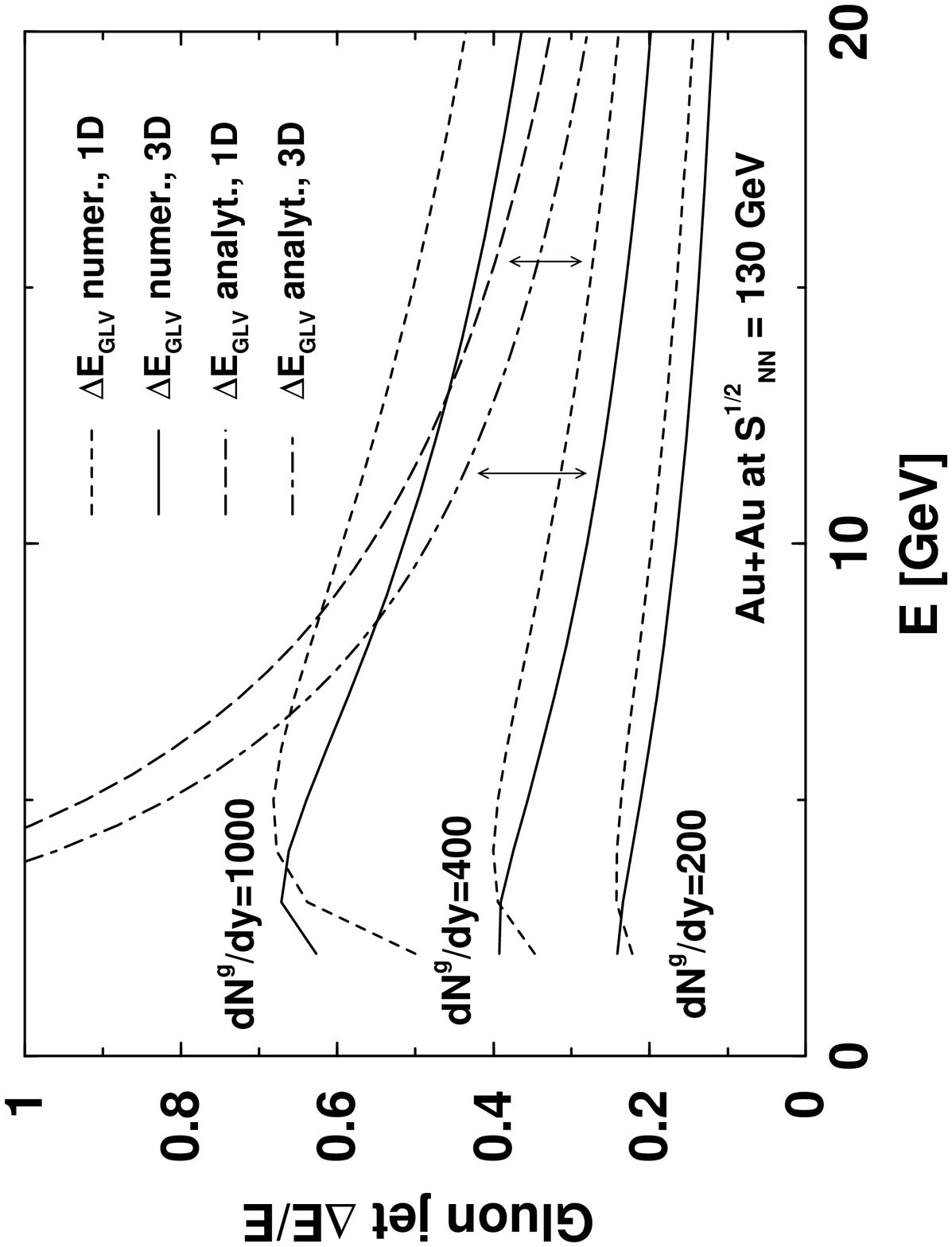} 
\end{center} 
\begin{center} 
\includegraphics{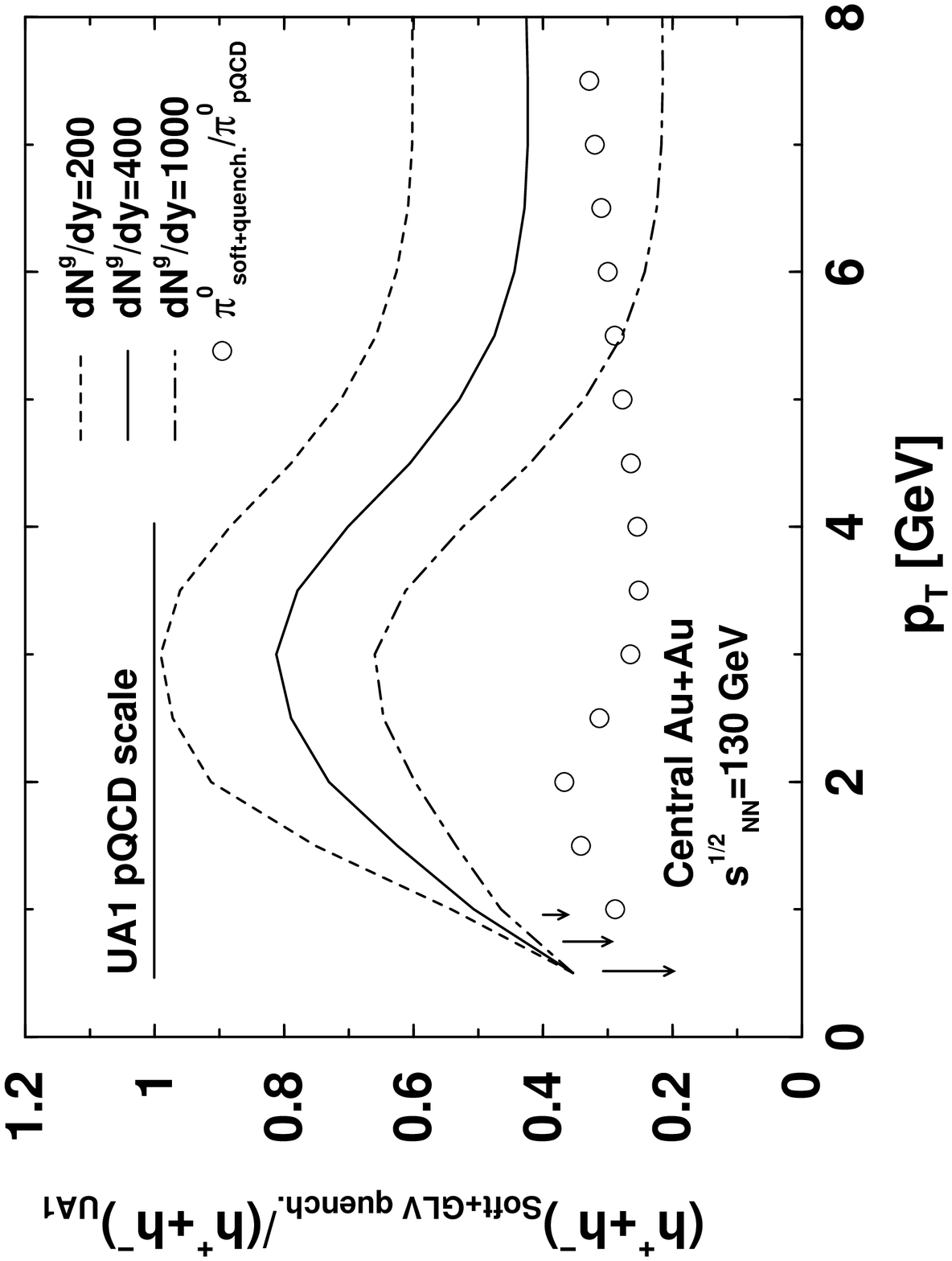} 
\vspace{-2.9cm} 
\end{center} 
 
\begin{center} 
\begin{minipage}[t]{15cm} 
         {\small {\bf FIGURE 1.}  In Fig. 1a the gluon jet fractional energy  
loss in the GLV formalism is shown for three  
initial gluon rapidity densities $dN^g/dy=200, 400, 1000$ for 
both 1D and 3D expansion. The asymptotic {\em analytic} approximation ( 
shown here for $dN^g/dy=400$)  
is inapplicable at RHIC.  
Fig. 1b shows the ratio of charged particles computed in the 
soft+GLV quenched model relative to the conventional PQCD that accounts for the 
UA1 data. Also shown is the greater quenching predicted for 
$\pi^0$ with $dN^g/dy=400$.}  
\end{minipage} 
\end{center} 
 
\noindent We have also studied the effects of a strong transverse 
flow in addition to the boost-invariant longitudinal Bjorken  
expansion~[13] and for the azimuthally  
averaged energy loss and found the simple result: 
\begin{equation} 
\Delta E^{(1)}_{\rm 3D} \approx 
\frac{\delta}{1+\delta v_\perp } \Delta E^{(1)}_{\rm 1D},\qquad 
z_{\rm max}=\delta R \;\;, 
\label{1d3dfac} 
\end{equation}  
where  $v_\perp$ is the average 
transverse expansion velocity and  $z_{\rm max}=\delta R$ is the  
duration of the energy loss phase ($\delta < 1/(1-v_\perp)$).  
While the azimuthal asymmetry of $\Delta E$ is sensitive  
to the details of the expansion, the mean is not.  
For   $v_\perp=0.6$ and $\beta = 1.5$  
$\Delta E^{(1)}_{\rm 3D} \simeq 80\% \Delta E^{(1)}_{\rm 1D}$. Gluon jet mean 
fractional energy loss for three typical plasma densities is shown in 
Fig. 1a. Comparison to the asymptotic result 
signifies the importance of finite 
kinematics. Non-abelian energy loss affects hadronization through rescaling 
the hadron momentum fractions $z_c$ and modifying the fragmentation functions 
in Eq.~(\ref{hcrossec})~[6]. We note that because of the  
GLV energy loss  
$\Delta E/E\propto \rho_g$ is very weakly energy dependent (see Fig. 1a), 
 the effect of energy loss fluctuations,  
can be absorbed as  a gluon density 
renormalization by a factor $\sim 2-2.5$~[14].

The PQCD fragmentation picture 
gives adequate description of  
particle multiplicities in $pp$ and $\bar{p}{p}$ collisions 
for $p_{\rm T} > 2-3$~GeV but  over-predicts   
soft hadrons due to its power law divergence.  
We use the the results of the soft string models that fit $pp$ and   
$\bar{p}p$ data to parametrize  by a simple 
exponential $p_{\rm T}$ behavior (in the $m \rightarrow 0$ limit)   
the low $p_{\rm T}$ region.  
If similar soft string dynamics drives $pp,\; pA$ and $AA$ 
collisions then most of the global observables with the exception of the 
baryon-antibaryon asymmetry can also be inferred from already existing data.  
The FNAL E-735 data~[15] on hadronic spectra in  $\bar{p}{p}$ 
at $\sqrt{s}=200$~GeV  suggests values for the particle ratios  that come very 
close to preliminary RHIC data (shown for comparison in parentheses). 
The $\bar{p}/\pi=0.07$ (0.08). For $K/\pi=0.11$ (0.15) we find  
$\sim 30\%$ deviation.  For pions and kaons one infers 
from charge and strangeness conservation that $\pi^+/\pi^-=1.0$ (1.0)  
and $K^+/K^- = 1.0$ (1.1).          
The ratio that cannot be fixed from data on net baryon-free $\bar{p}p$  
collisions  is $\bar{p}/p$. However,  
it can be predicted by baryon transport (Regge) theory, i.e.  
${dN^B}/{dy}= Z \beta {\cosh ((1-\alpha_B(0))y)} 
/{\sinh((1-\alpha_B(0))Y_{\rm max})}$ 
with a coupling $\beta = {\cal O}(1)$. 
In the baryon Junction picture  $\alpha_B(0) \simeq 1/2$~[11]. 
For $Au+Au$ collisions at RHIC at $\sqrt{s}=130$~AGeV ($Y_{\rm max}=5.4$), 
the prediction $dN^B/dy \sim  
 dN^p/dy-dN^{\bar p}/dy\simeq 11$ is close to the preliminary data.  
The predicted Junction trajectory slope, 
$\alpha_J^\prime\approx \alpha_R^\prime/3$,  
implies that the {\em effective} string tension $\kappa^\prime$ 
is  three times higher than  
$\kappa=1/(2\pi\alpha_R^\prime)\approx 1$~GeV/fm. 
This  naturally leads to  $\langle p_{\rm T}^2\rangle_J\approx  
3 \langle p_{\rm T}^2 \rangle_R$, i.e.  
 $T_0^p \simeq \sqrt{3} T_0^\pi$.   
Given the pion mean inverse slope, $T_0^\pi \simeq 220$~MeV,  it 
{\em follows} that the proton and antiproton slopes should be   
$T_0^p \simeq 400$~MeV for the junction (high $p_{\rm T}$) 
component of those yields.

\newpage 
 
\vspace*{7.5cm} 
\begin{center} 
\includegraphics{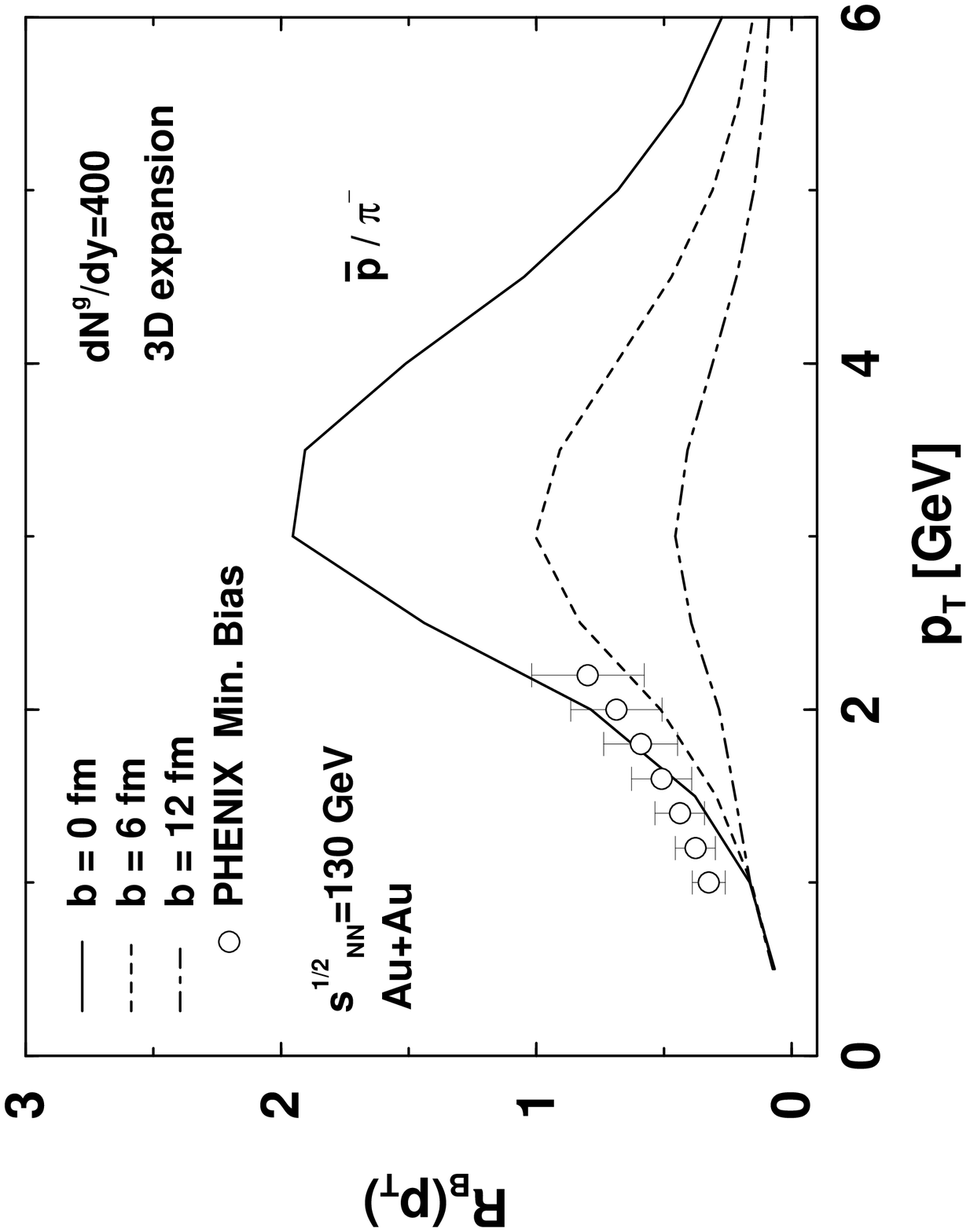} 
\end{center} 
\begin{center} 
\includegraphics{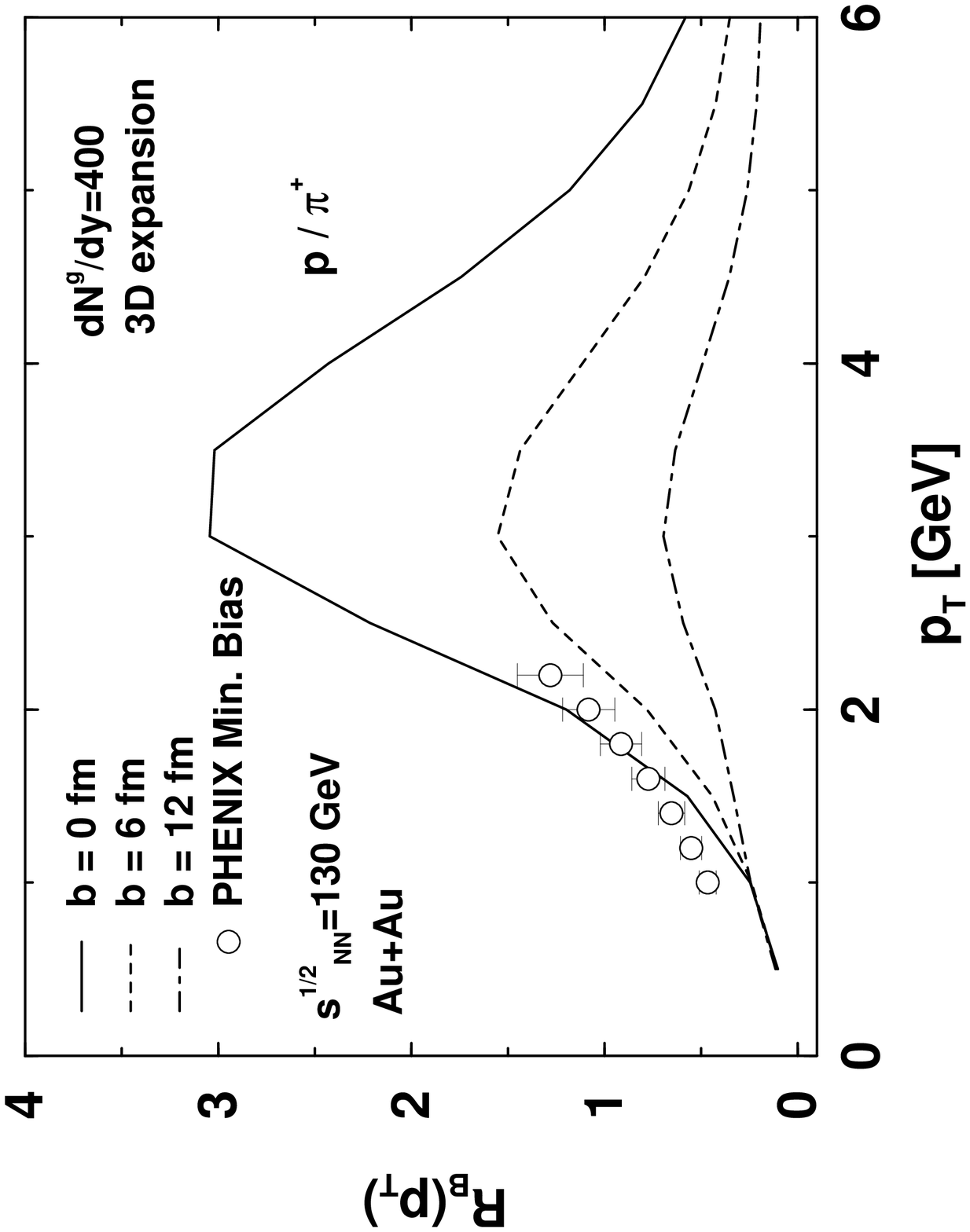} 
\vspace{-2.9cm} 
\end{center} 
 
\begin{center} 
\begin{minipage}[t]{15cm} 
         {\small {\bf FIGURE 2.}  The $\bar{p}/\pi^-$ and $p/\pi^+$ 
ratios are shown  in Figs. 2a, 2b as a function of $p_{\rm T}$ for  
3 different impact parameters ${\bf b} = 0, 6, 12$~fm.    
The ratio of out fits to preliminary  minimum bias  PHENIX data is  
shown for comparison.} 
\end{minipage} 
\end{center}

Published RHIC data~[4]  constrain the  
charged particle multiplicity  at midrapidity to  
$dN^h/dy({\bf b}=0) \simeq 600$.   
Thus with one parameter, the pion mean inverse slope  
(the $T_0^K=275$~MeV is irrelevant for the study of the $\bar{p}/\pi^-$  
and $p/\pi^+$ ratios), 
the above theoretical analysis based on the string and baryon Junctions  
picture results in a solvable set of equations for the parameters in the  
phenomenological soft component of the hadronic spectrum:   
\begin{equation} 
\frac{dN_s^{h} ({\bf b})}{dyd^2{\bf p}_{\rm T}} =  
\sum\limits_{\alpha=\pi,K,p,\cdots} 
\frac{dn^\alpha({\bf b})}{dy} \times \frac{e^{-p_{\rm T}/T_0^\alpha({\bf b})}} 
{2\pi T^{\alpha\;2}_0({\bf b})} 
\; . \label{flavslope} 
\end{equation} 
This component is assumed to scale with the number of participants, 
although the Junction component may scale  faster~[11]. 
Eqs.~(\ref{hcrossec},\ref{flavslope}) give the   
hadron inclusive multiplicity   
\beq 
\frac{dN^{h} ({\bf b})}{dyd^2{\bf p}_{\rm T}} = 
\frac{dN^{h}_s ({\bf b})}{dyd^2{\bf p}_{\rm T}}  
+  
T_{AA}({\bf b})  
\frac{d\sigma^{pp}} {dyd^2{\bf p}_{\rm T}}\;, 
\eeq{hadtot} 
where $T_{AA}({\bf b})$ is the Glauber profile density per unit area. 
The power law  PQCD component is limited to 
the range $p_{\rm T}>2$~GeV. 
If the Junction and Junction loop components 
scales nonlinearly with $N_{part}(b)$, 
as in HIJING/$B\bar{B}$, then the rate of change 
of the  maximum of $R_B$ 
with increasing ${\bf b}$ would change correspondingly. 
 
\vspace*{4mm} 
 
\begin{center} 
{\large  \bf CONCLUSIONS} 
\end{center} 
 
Fig.~1b shows the ratio of the charged particle multiplicities computed 
in our model 
relative to unquenched scaled PQCD (which fits to the UA1 data~[16]).  
The high $p_{\rm T}$ 
tails show approximately constant suppression driven by the initial gluon 
density. Similar computation is shown for neutral pions for  
mean energy loss driven calculation with $dN^g/dy=400$. In the intermediate 
$p_{\rm T} =2-5$~GeV region the factor of $~2$ difference in the  
{\em apparent} suppression arises 
from the anomalously large baryon and antibaryon component 
predicted by the Junction picture.   
 
In going from central to peripheral collisions the smaller partonic densities 
result in  larger typical sizes of strings and Junctions and somewhat smaller   
mean inverse slopes of the resulting soft hadrons. This effect is taken  
into account by introducing a difference $\Delta T_0^{\pi}=-10$~MeV, 
$\Delta  T_0^{K}=-25$~MeV and  $\Delta T_0^{p} =- 50$~MeV  
between central ${\bf b} =0$~fm and peripheral 
${\bf b} =12$~fm  collisions. The major distinction between various 
centralities, however, is the different strength of jet quenching due to the  
different lengths of the path that the jet travels inside the plasma and 
the different density of the medium. Thus in peripheral collisions the  
baryon/meson ratio is expected to stay below unity similar to the $\bar{p}p$ 
case. Figs. 2a and 2b  show  the hybrid two component model predictions for 
the $\bar{p}/\pi^-$ and $p/\pi^+$ ratios as a function of transverse momentum 
and centrality. While the precise values $R_{B \; {\rm max}}$  
of the  maximum baryon and antibaryon excess  
are sensitive to the uncertainties in the estimates 
of the plasma density and the mean inverse slopes, the characteristic 
$p_{\rm T}$ and ${\bf b}$ behavior is not. In Figs. 2a and 2b we have 
included the ratio of our exponential fits in a narrow $p_{\rm T}=1-2.1$~GeV 
window to the preliminary PHENIX minimum bias data~[10]. The error 
bars reflect the one $\sigma$ uncertainties in the slope and intercept 
parameters.

Unlike hydrodynamics the dual soft+hard hybrid model predicts that the  
baryon and antibaryon anomaly is in fact limited to finite $p_{\rm}$  
window and the baryon to meson/ratios eventually converge to the PQCD  
dominated base at high $p_{\rm T}$. It also predicts the systematic  
decrease of the  predicted ratios with impact parameter which can be  
tested at RHIC. With higher statistics the dynamics of the baryon/meson 
ratios as a function of centrality may  fact be probed in the  
current experimentally accessible $p_{\rm T}$ range.

\vspace*{4mm} 
 
\begin{center} 
{\large  \bf REFERENCES} 
\end{center}

{\small 
\noindent 1. \   Quark Matter 2001,  
http://www.rhic.bnl.gov/qm2001/program.html 
\\  
2. \  David,~G, [PHENIX Collaboration],  nucl-ex/0105014;  
Zajc,~W.A., [PHENIX Collaboration], nucl-ex/0106001. 
\\ 
3. \  Ackermann,~K.H. {\it et al.}, [STAR Collaboration], 
Phys. Rev. Lett. 86, 402 (2001),  
[nucl-ex/0009011]; 
Snellings,~R.J., \\ 
\hspace*{4mm} [STAR Collaboration], 
nucl-ex/0104006. 
\\ 
4. \  Back,~B.B. {\it et al.},  [PHOBOS Collaboration], 
Phys. Rev. Lett. 85, 3100 (2000), 
[hep-ex/0007036]; 
Adcox,~K. \\ 
\hspace*{4mm} {\it et al.},  [PHENIX Collaboration], 
Phys. Rev. Lett. 86, 3500 (2001),  
[nucl-ex/0012008]. 
\\ 
5. \  Wang,~X.-N. and Gyulassy,~M., 
Phys. Rev. Lett. 86, 3496 (2001), [nucl-th/0008014]. 
\\ 
6. \  Gyulassy,~M., Vitev,~I and Wang,~X.-N., 
Phys. Rev. Lett. 86, 2537 (2001);  
L\'evai,~P.  {\it  et al.}, nucl-th/0104035. 
\\ 
7. \  Gyulassy,~M., L\'evai,~P., Vitev,~I., 
Phys. Rev. Lett. 85, 5535, (2000);  
Nucl. Phys.  B594, 371 (2001).  
\\  
8. \  Aggarwal,~M.M. {\it et al.},  [WA98 Collaboration], 
Phys. Rev. Lett. 81, 4087 (1998); Erratum-{\it ibid}.  
84, 578 (2000). 
\\ 
9.  \  Wang,~X.-N., 
Phys. Rev. Lett. 81, 2655 (1998), 
[hep-ph/9804384]. 
\\ 
10. \  Velkovska,~J., [PHENIX Collaboration], nucl-ex/0105012.  
\\ 
11. \  Rossi,~G.C. and Veneziano,~G., 
  Nucl. Phys. B123, 507  (1977);  
   Kharzeev,~D., Phys. Lett. B378,  238 (1996); \\  
 \hspace*{5.5mm}  Kopeliovich,~B.Z. and Zakharov,~B.G., 
   Z. Phys. C43, 241 (1989); 
  Phys. Lett. B381, 325 (1996);    
   Vance,~S.E.,  \\ 
 \hspace*{5.5mm}  Gyulassy,~M. and Wang,~X.-N., 
         Phys. Lett. B443, 45 (1998); 
        Vance,~S.E. and Gyulassy,~M., 
        Phys. Rev.  Lett. \\  \hspace*{5.5mm} 83,  1735 (1999). 
\\  
12. \  Xu,~N. and  Kaneta,~M., [STAR Collaboration], 
 nucl-ex/0104021;  
Harris,~J.,   ``Results from STAR",  \\  \hspace*{5.5mm}
http://www.rhic.bnl.gov/qm2001/program.html 
\\ 
13. \   Vitev,~I. and Gyulassy,~M., nucl-th/0104066;   
Vitev~I. and Gyulassy~M., work in progress. 
\\ 
14. \  Gyulassy,~M., Levai,~P. and Vitev,~I., in preparation. 
\\  
15. \  Alexopoulos,~T. {\it et al.}, 
[E735 Collaboration], 
Phys. Rev. D48 984 (1993). 
\\ 
16. \  Albajar,~C {\it et al}, 
[UA1 Collaboration], Nucl. Phys. B335, 261 (1990). 
\\
}
 
\vfill\eject 
\end{document}